\documentclass[11pt]{article}
\usepackage{amsmath}
\usepackage{amsthm}
\usepackage{graphicx}
\usepackage{multirow}
\usepackage{booktabs}
\usepackage[numbers]{natbib}
\usepackage{bm}
\usepackage{geometry}
\usepackage{setspace}
\usepackage{pdfpages}
\newcommand\independent{\protect\mathpalette{\protect\independenT}{\perp}}
\def\independenT#1#2{\mathrel{\rlap{$#1#2$}\mkern2mu{#1#2}}}
\DeclareMathOperator{\var}{var}
\DeclareMathOperator{\SPE}{SPE}
\DeclareMathOperator{\expit}{expit}
\allowdisplaybreaks[4]
\newtheorem{theorem}{Theorem}
\newtheorem{lemma}{Lemma}
\newtheorem{assumption}{Assumption}
\newtheorem{corollary}{Corollary}

\title{\bf Separable pathway effects of semi-competing risks using multi-state models}

\author{Yuhao Deng, Yi Wang, Xiang Zhan and Xiao-Hua Zhou}

\begin{document}

\maketitle

\abstract{Semi-competing risks refer to the phenomenon where a primary event (such as mortality) can ``censor'' an intermediate event (such as relapse of a disease), but not vice versa. Under the multi-state model, the primary event consists of two specific types: the direct outcome event and an indirect outcome event developed from intermediate events. Within this framework, we show that the total treatment effect on the cumulative incidence of the primary event can be decomposed into three separable pathway effects, capturing treatment effects on population-level transition rates between states. We next propose two estimators for the counterfactual cumulative incidences of the primary event under hypothetical treatment components. One estimator is given by the generalized Nelson--Aalen estimator with inverse probability weighting under covariates isolation, and the other is given based on the efficient influence function. The asymptotic normality of these estimators is established. The first estimator only involves a propensity score model and avoid modeling the cause-specific hazards. The second estimator has robustness against the misspecification of submodels. As an illustration of its potential usefulness, the proposed method is applied to compare effects of different allogeneic stem cell transplantation types on overall survival after transplantation. \par
}

\maketitle

\section{Introduction} \label{sec1}

In many clinical trials focusing on time-to-event outcomes, there may be a primary (terminal) event and an intermediate (non-terminal) event. The terminal event can ``censor'' the non-terminal event, but not vice versa. This phenomenon is referred to as semi-competing risks \cite{fine2001semi, xu2010statistical, haneuse2016semi, zhang2024marginal}. In the presence of semi-competing risks, some individuals have observations on both intermediate and primary events, while others only have observations on primary events. Although all individuals can experience primary events, the hazard of developing primary events may be different between those individuals with intermediate events and those without intermediate events. That is, the intermediate event can modify the hazard of the terminal event. The intermediate event and terminal event are not independent, so conventional intention-to-treat analysis would not recover the pure treatment effect on the terminal event.

Our work is motivated by a study of allogeneic stem cell transplantation to treat acute lymphoblastic leukemia \cite{ma2021an}. Mortality is a terminal event and relapse of leukemia is an intermediate event. Clinical researches have indicated that haploidentical transplantation leads to a higher transplant-related mortality but lower relapse rate compared with human leukocyte antigens matched transplantation \cite{chang2020haploidentical}. Throughout this article we consider the terminal event as the primary event, whereas the treatment effect on the intermediate event is related to the survivor average causal effect \cite{xu2022bayesian, nevo2022causal}. The hypothetical incidence of the primary event by appropriately adjusting the intermediate event is of primary interest. New estimands and inference methods to deal with semi-competing risks under the causal inference framework are desired.

Typically there are two approaches to infer the pure treatment effect on the primary event by adjusting intermediate events. The first approach is to restrict the target population to the principal stratum on which no intermediate events would occur no matter which treatment is applied \cite{gao2023defining}. Since the target population is unobservable, it is difficult for policy making. Moreover, principal stratification only focuses on the treatment effect in a subpopulation, and thus cannot identify the effect modified by intermediate events. Another approach to study the causal mechanism of semi-competing risks is mediation analysis. Huang \cite{huang2021causal} defined the natural direct and indirect effects on the cumulative hazard of the primary event and illustrated the nonparametric identifiability of these effects. However, this natural direct effect essentially does not distinguish an interaction effect resulted by the competing nature of primary outcome events and intermediate events \cite{vanderweele2013three}. In addition, how to incorporate covariates in the analysis requires further clarification.

Both the principal stratification and mediation analysis approaches require untestable principal ignorability or sequential ignorability for identification. To avoid untestable assumptions in principle, an interventionist approach attempts to decompose the initial treatment into two components as an alternative to mediation analysis \cite{robins2010alternative}. The separable effects framework provides easier interpretation by avoiding cross-world assumptions and targets relevant clinical questions for future experiments \cite{stensrud2021discussion}. For longitudinal or time-to-event data, we have an event process for the intermediate event and another event process for the terminal (primary) event. Counterfactual processes are generated when the treatment components are not all the same \cite{martinussen2021estimation}. Stensrud et al.\cite{stensrud2021generalized, stensrud2022separable} studied the separable effects in the presence of competing events. They identified the partial isolation condition under which the counterfactual incidence of the primary event is identifiable. The core assumption is the dismissible treatment components condition, saying that each treatment can only have a direct effect on one event conditional on the history. Then they proposed a regression estimator and a weighting estimator. However, the asymptotic properties of the estimators are hard to explore. Later, Breum et al. \cite{breum2024estimation} proposed parametric estimation based on influence functions for semi-competing risks.

In the semi-competing risks context, the terminal event can be developed either following treatment or following the intermediate event. Therefore, the treatment effect on the terminal event relies on the heterogeneous cause-specific hazards of terminal events. The direct outcome event following treatment and the indirect outcome event following the intermediate event can be regarded as different events as they may have different mechanisms of occurrence. In the motivating data, transplant-related mortality (non-relapse mortality) is caused by delayed immune reconstitution, whereas the relapse-related mortality is caused by the decline of normal leukocyte. The total treatment effect should be decomposed into three separable effects, corresponding to the direct outcome event, intermediate event and indirect outcome event, respectively. It is especially challenging to identify the separable effect on the indirect outcome event because the time origin of the transition from intermediate events to indirect outcome events is not naturally aligned. Furthermore, we should try to avoid modeling the cause-specific hazards parametrically since it is very difficult to specify the model correctly for each competing event.

In this paper, we study the causal mechanism of semi-competing risks under the potential outcomes framework based on multi-state models. We divide the primary outcome events into two states: a direct outcome event state and an indirect outcome event state. The target estimand is defined as the counterfactual cumulative incidence of the terminal event (including the direct outcome event and indirect outcome event) by manipulating the treatment components. The contribution of our paper is fourfold. First, we show that population-level hazards of direct outcome event, intermediate event and indirect outcome event are identifiable by inverse probability weighting under a condition of covariates isolation, inspiring a weighted Nelson--Aalen estimator. Second, we derive counterfactual cumulative incidence functions of the terminal event by intervening on different treatment components, and then define separable pathway effects based on these counterfactual cumulative incidences. We show the identifiability of these separable pathway effects under appropriate assumptions. Third, we propose a generalized Nelson--Aalen estimator by inverse probabiliy weighting (GNAIPW) under covariates isolation, and propose an efficient influence function (EIF) based estimator in the general case with conditional dismissible treatment components. Fourth, we establish asymptotic properties of the estimated counterfactual cumulative incidence functions. Hypothesis tests for the treatment effect on transition hazards or separable pathway effects are then constructed using logrank statistics or influence functions. Through a real-data application on allogeneic stem cell transplantation, we illustrate the usefulness of the proposed methods to detect the mechanism of treatment effects from a total effect.

The remainder of this paper is organized as follows. Section \ref{sec2} introduces the framework and notations of semi-competing risks, and then shows the identification of the counterfactual cumulative incidences under appropriate assumptions. We first consider a condition of covariates isolation under which the population-level transition hazards are identifiable, and the counterfactual cumulative incidences are derived from the population-level transition hazards. Then we consider the general conditions in order that the separable pathway effects are identifiable with relaxed assumptions. Under the covariates isolation, we propose a generalized Nelson--Aalen estimator for the population-level transition hazards by inverse probability weighting in Section \ref{sec3}. When covariates isolation fails, we propose an estimator for the counterfactual cumulative incidence based on efficient influence functions in Section \ref{sec4}. The asymptotic properties for the estimated counterfactual cumulative incidences are established. We conduct simulation studies to assess the performance of the proposed estimators in Section \ref{sec6}. The proposed method is applied to allogeneic stem cell transplantation data in Section \ref{sec7}. Finally this paper ends with a discussion in Section \ref{sec8}.

\section{Identification of Treatment Effects} \label{sec2}

\subsection{Framework and notations}

Consider an experiment with a binary treatment $A\in\{0,1\}$ conducted on $n$ units. Let $T^{a}$ be the potential failure time of the primary event and $R^a$ be the potential failure time of the intermediate event when the treatment is set at $A = a$. We define $R^a > \tau$ if no intermediate event happens before the primary event, where $\tau$ is the end time of the experiment. Let $C^a$ be the potential censoring time, let $\delta_T^a = I(T^a\le C^a)$, $\delta_R^a = I(R^a\le C^a)$ be potential event indicators, and let $\widetilde{T}^a = \min(T^a,C^a)$, $\widetilde{R}^a = \min(R^a,C^a)$ be the potential event times.
Throughout this article, we assume the stable unite treatment value assumption (SUTVA) that all individuals are independent of each other. We assume consistency for the potential event times and indicators as follows.
\begin{assumption}[Consistency] \label{consis}
$(\widetilde{T},\widetilde{R},\delta_T,\delta_R) = (\widetilde{T}^A,\widetilde{R}^A,\delta_T^A,\delta_R^A).$
\end{assumption}

In addition to these time-to-event variables, some baseline covariates $X$ may also be collected in the experiment. Our sample includes $n$ independent and identically distributed copies of $(A,\widetilde{T},\widetilde{R},\delta_T,\delta_R,X)$, indexed by subscripts $i\in\{1,\ldots,n\}$. We assume the treatment assignment mechanism is ignorable.

\begin{assumption}[Ignorability] \label{ign}
$(T^a, R^a, C^a) \independent A \mid X$, for $a=0,1$.
\end{assumption}

This data generating process can be understood using a multi-state model. The primary event is decomposed into two states: direct outcome event following treatment and indirect outcome event following intermediate event. The direct outcome event and intermediate event are a pair of competing events. The time origin of transition from the initial status (State 0) to the direct outcome event (State 1) or intermediate event (State 2) is 0, and the time origin of transition from intermediate event to indirect outcome event (State 3) is $R^a$. Denote the counterfactual cumulative incidences for these three states by $F_1^a(t)=P(T^a\le t, T^a<R^a)$, $F_2^a(t)=P(R^a\le t, R^a\le T^a)$ and $F_3^a(t)=P(T^a\le t, R^a\le T^a)$. Then, the counterfactual cumulative incidence of the primary event $F^a(t)=P(T^a\le t)=F_1^a(t)+F_3^a(t)$.

Suppose the treatment $A$ can be decomposed into three components $(A_1,A_2,A_3)$, where $A_j$ only has an effect on the hazard of transiting to State $j$. In a real-world trial, the actual treatment is equal to separable treatment components, $A=A_1=A_2=A_3$. In a hypothetical world, these treatment components can take different values. Hereafter we suppose $a=(a_1,a_2,a_3)$ is a three-dimensional vector. A scalar $a$ represents $(a,a,a)$. Under the treatment components combination $(a_1,a_2,a_3)$, let $F^{(a_1,a_2,a_3)}(t)$ be the counterfactual cumulative incidence of the terminal event. The total treatment effect is decomposed as
\begin{align*}
F^{(1,1,1)}(t) - F^{(0,0,0)}(t)  &=\{F^{(1,0,0)}(t) - F^{(0,0,0)}(t)\} + \{F^{(1,1,1)}(t) - F^{(1,0,0)}(t)\}  \\
&:= \SPE_{0\to1}(t;0,0) + \SPE_{0\to3}(t;1) \notag \\
&= \{F^{(1,0,0)}(t) - F^{(0,0,0)}(t)\} + \{F^{(1,1,0)}(t) - F^{(1,0,0)}(t)\} \notag \\
&\qquad + \{F^{(1,1,1)}(t) - F^{(1,1,0)}(t)\}\\
&:= \SPE_{0\to1}(t;0,0) + \SPE_{0\to2}(t;1,0) + \SPE_{2\to3}(t;1,1),
\end{align*}
with $\SPE_{0\to1}(t;0,0)$ representing the direct effect on the primary event, $\SPE_{0\to3}(t;1)$ representing the indirect effect via the intermediate event. $\SPE_{0\to3}(t;1)$ can be further divided into two parts, with $\SPE_{0\to2}(t;1,0)$ representing the effect of modifying the risk of intermediate events via the transition from State 0 to 2, and $\SPE_{2\to3}(t;1,1)$ representing the interaction effect of treatment and intermediate event via the transition from State 2 to 3.

Denote $d\Lambda_j^a(t\mid\mathcal{F}(t))$ as the counterfactual hazard of the transition to State $j$ at time $t$ given a set of history $\mathcal{F}(t)$ up to $t$ under the hypothetical treatment $a=(a_1,a_2,a_3)$, with
\begin{align}
d\Lambda_1^{a}(t\mid\mathcal{F}(t)) &:= P(t\le T^a < t+dt, T^a<R^a \mid T^a\ge t, R^a\ge t, \mathcal{F}(t)), \\
d\Lambda_2^{a}(t\mid\mathcal{F}(t)) &:= P(t\le R^a < t+dt, R^a\le T^a \mid T^a\ge t, R^a\ge t, \mathcal{F}(t)), \\
d\Lambda_3^{a}(t\mid\mathcal{F}(t)) &:= P(t\le T^a < t+dt, R^a\le t \mid T^a\ge t, R^a\le t, \mathcal{F}(t)),
\end{align}
where $dt\to0$ and $r \le t$. Since the hazards involves cross-world quantities, they are not directly estimable using observed data. To separate the effects of each treatment components, we describe some scenarios of isolation in the next sections.

\subsection{Covariates isolation}

\begin{figure}
\centering
\includegraphics[width=0.95\textwidth]{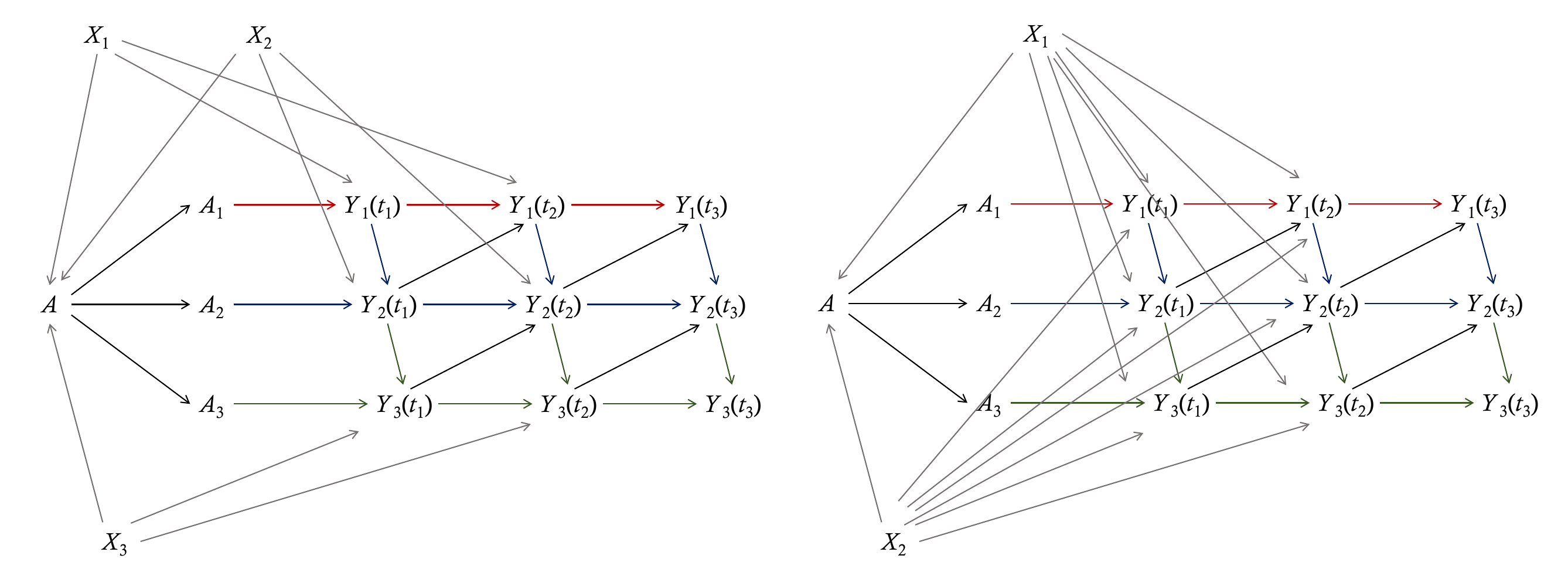}
\caption{Extended directed acyclic graph of semi-competing risks. The direct outcome event, intermediate event, indirect outcome event have indicator processes $Y_1(t)$, $Y_2(t)$ and $Y_3(t)$. Covariates isolation (left) and general case (right).} \label{edag}
\end{figure}

To better understand the mechanism of event processes, we discretize the event processes as indicators at discrete time points. First we assume that the effects of covariates are isolated on different events. The left panel of Figure \ref{edag} shows the extended directed acyclic graph of the semi-competing risks with isolated covariates effects. The treatment $A$ has three components $A_1$, $A_2$ and $A_3$. In the graph, $X_1$ only has an effect on the direct outcome event $Y_1(t)$, $X_2$ only has an effect on the intermediate event $Y_2(t)$, and $X_3$ only has an effect on the indirect outcome event $Y_3(t)$. At any time $t$, we monitor $Y_1(t)$, $Y_2(t)$ and $Y_3(t)$ in turn. If the direct outcome event does not happen, then there is a risk that the intermediate event happens. If the intermediate event happens, then the direct outcome event can never happen and there is a risk that the indirect outcome event happens. If the indirect outcome event happens, then the direct outcome event can never happen. Conditional on the event history $(Y_1(s),Y_2(s),Y_3(s): s<t)$, the paths from $(A_2,A_3)$ to $Y_1(t)$ are blocked, the paths from $(A_1,A_3)$ to $Y_2(t)$ are blocked, and the paths from $(A_1,A_2)$ to $Y_3(t)$ are blocked.

To account for censoring, we assume the censoring time is independent of the potential terminal and intermediate events. The cenoring time can rely on the treatment group, but should not have common causes with the terminal and intermediate events. We also require that the censoring time is large enough.

\begin{assumption}[Random censoring] \label{ran}
$C^a \independent (T^a, R^a)$, for $a=0,1$.
\end{assumption}
\begin{assumption}[Positivity] \label{pos}
$c<P(A=a \mid X)<1-c$ for a constant $c>0$,  $P(T^a>\tau, C^a>\tau \mid A=a)>0$ for any $0\le t\le\tau$ and $a=0,1$.
\end{assumption}

Under the covariates isolation, we let $d\Lambda_1^a(t)$ and $d\Lambda_2^a(t)$ be the hazard of the direct outcome event and intermediate event at time $t$ under the treatment combination $a=(a_1,a_2,a_3)$. We let $d\Lambda_3^a(t;r)$ be the hazard of the indirect outcome event at time $t$ given that the intermediate event happens at $r \leq t$. We formalize the assumption on the way treatment components exert effects on event processes.

\begin{assumption}[Dismissible components] \label{dis}
$d\Lambda_1^{(a_1,a_2,a_3)}(t) = d\Lambda_1^{a_1}(t)$,
$d\Lambda_2^{(a_1,a_2,a_3)}(t) = d\Lambda_2^{a_2}(t)$,
$d\Lambda_3^{(a_1,a_2,a_3)}(t;r) = d\Lambda_3^{a_3}(t;r)$,
with $dt\to0$.
\end{assumption}

Assumption \ref{dis} means that the population-level counterfactual hazard $d\Lambda_j^{(a_1,a_2,a_3)}(\cdot)$ only relies on the treatment component $a_j$ rather than covariates. This assumption holds if three sets of covariates have isolated effects on these three hazards respectively. For example, in stem cell transplantation, treatment-related mortality (due to low immunity), relapse (due to minimum residual disease) and relapse-related mortality (due to abnormal leukocyte) has different biological mechanisms. Different sets of covariates are found to moderate the risk of each event: age affects the risk of relapse-related mortality, while diagnosis type affects the risks of relapse.

When $a_1=a_2=a_3$, Assumption \ref{dis} is naturally satisfied because no hypothetical worlds are involved. The plausibility of this assumption when $a_1$, $a_2$ and $a_3$ are not all equal can be assessed by future experiments if the treatment components are discovered \cite{stensrud2022separable}. In fact, Assumption \ref{dis} can also hold if the effects of covariates are not fully isolated. A theoretical example of dismissible components is the additive hazards model; see Supplementary Material A.2.

\begin{theorem}\label{thm1}
Under Assumptions \ref{consis}--\ref{dis}, the hazards $d\Lambda_j^a(\cdot)$ ($j=1,2,3$) are identifiable. The counterfactual cumulative incidences are identifiable,
\begin{align}
F_1^{(a_1,a_2,a_3)}(t) &= \int_0^t\exp\{-\Lambda_1^{a_1}(s)-\Lambda_2^{a_2}(s)\}d\Lambda_1^{a_1}(s), \\
F_2^{(a_1,a_2,a_3)}(t) &= \int_0^t\exp\{-\Lambda_1^{a_1}(s)-\Lambda_2^{a_2}(s)\}d\Lambda_2^{a_2}(s), \\
F_3^{(a_1,a_2,a_3)}(t) &= F_2^{(a_1,a_2,a_3)}(t) - \int_0^t\exp\{-\Lambda_1^{a_1}(s)-\Lambda_2^{a_2}(s)-\Lambda_3^{a_3}(t;s)\}d\Lambda_2^{a_2}(s).
\end{align}
\end{theorem}

Proof of Theorem \ref{thm1} is given in Supplementary Material A.1.

\subsection{General case}

When the covariates have interaction effects on events, the covariates isolation is likely to fail. Suppose that there are no post-treatment covariates. The right panel of Figure \ref{edag} illustrates the general case of separable effects. Conditional on all baseline covariates $X$ and the history $(Y_1(s),Y_2(s),Y_3(s): s<t)$, the paths from $(A_2,A_3)$ to $Y_1(t)$ are blocked, the paths from $(A_1,A_3)$ to $Y_2(t)$ are blocked, and the paths from $(A_1,A_2)$ to $Y_3(t)$ are blocked. Especially, for the indirect outcome event, the event history is the occurrence time of the intermediate event. Compared with covariates isolation, the general case allows measured common confounding for different events. We formalize the random censoring and dismissible components as follows.

\begin{assumption}[Conditional random censoring] \label{ran_pi}
$C^a \independent (T^a, R^a) \mid X$, for $a=0,1$.
\end{assumption}
\begin{assumption}[Conditional positivity] \label{pos_pi}
$c<P(A=a \mid X)<1-c$ for a constant $c>0$,  $P(T^a>\tau, C^a>\tau \mid A=a, X)>0$ for any $0\le t\le\tau$ and $a=0,1$.
\end{assumption}
\begin{assumption}[Conditional dismissible components] \label{dis_pi}
$d\Lambda_1^{(a_1,a_2,a_3)}(t;x) = d\Lambda_1^{a_1}(t;x)$,
$d\Lambda_2^{(a_1,a_2,a_3)}(t;x) = d\Lambda_2^{a_2}(t;x)$,
$d\Lambda_3^{(a_1,a_2,a_3)}(t;r,x) = d\Lambda_3^{a_3}(t;r,x)$,
with $dt\to0$.
\end{assumption}

The dismissible components assumption hold conditional on baseline covariates. Under Assumptions \ref{consis}, \ref{ign}, \ref{ran_pi}--\ref{dis_pi}, the conditional hazards $d\Lambda_j^a(\cdot;x)$ ($j=1,2,3$) and the counterfactual cumulative incidences are identifiable. The identification result is straightforward as we can construct covariates isolation if we conduct the analysis in each level of the common risk factors for events and then take average over the distribution of baseline covariates, although at a cost of efficiency loss. With the conditional counterfactual cumulative incidence $F^a(t;x)$ identified, we integrate covariates out to get $F^a(t)$.


However, such a modification over covariates isolation brings challenges in estimating the counterfactual cumulative incidences. Covariates isolation allows estimation by weighting without involving survival time models. The direct outcome event and intermediate event share the same at-risk set due to competing risks. By appropriately weighting, the covariates distribution between the treated and control groups is balanced in the at-risk sets for the direct outcome event and intermediate event at the same time. However, since the at-risk set of the indirect outcome event is a subset of units with a history of intermediate event influenced by treatment components $(A_1,A_2)$, weighting cannot balance the covariates between the treated and control groups in the at-risk set of the indirect outcome event. Therefore, we need to employ survival time models to identify the counterfactual cumulative incidences if covariates isolation fails.

If there are time-varying covariates, additional assumptions are required for identification; see Supplementary Material G. Stensrud et al. \cite{stensrud2021generalized} identified the partial isolation condition for competing events in longitudinal studies in order that the counterfactual incidences are identifiable. The covariates should be divided into some parts, and each treatment component can only have a direct effect on one of the covariates parts.

\section{Estimation under covariates isolation} \label{sec3}

In this section, we apply weighted Nelson--Aalen estimators to estimate the separable pathway effects under covariates isolation. Since the hazards may depend on baseline covariates, we use inverse probability weighting to create a pseudo sample with balanced covariates between treatment groups. Let $w_i(a_j)=I\{A_i=a_j\}/P(A_i=a_j \mid X_i), i=1,\ldots,n,j=1,2,3$ denote the inverse of propensity score. Define weighted counting processes, at-risk processes and residuals with respect to $d\Lambda_1^{a_1}(t)$ and $d\Lambda_2^{a_2}(t)$ as follows:
\begin{align*}
N_{1}(t;a_1) &= \sum_{i=1}^{n}w_i(a_1)I\{\widetilde{T}_i\le t, \widetilde{R}_i>t, \delta^T_i=1\}, \
N_{2}(t;a_2) = \sum_{i=1}^{n}w_i(a_2)I\{\widetilde{R}_i\le t, \widetilde{T}_i\ge t, \delta^R_i=1\}, \\
Y_j(t;a_j) &= \sum_{i=1}^{n}w_i(a_j)I\{\widetilde{T}_i\ge t, \widetilde{R}_i\ge t\}, \
Y_j^w(t;a_j) = \sum_{i=1}^{n}w_i(a_j)^2I\{\widetilde{T}_i\ge t, \widetilde{R}_i\ge t\}, \\
M_j(t;a_j) &= \int_0^t \left\{dN_j(s;a_j)-Y_j(s;a_j)d\Lambda_j^{a_j}(s)\right\}, \ j=1,2.
\end{align*}
At first sight, it may seem straightforward to define weighted counting process $N_3(t;r,a_3)$, at-risk process $Y_3(t;r,a_3)$ and residual $M_3(t;r,a_3)$ with respect to $d\Lambda_3^{a_3}(t;r)$ in a similar manner as those aforementioned quantities for $d\Lambda_1^{a_1}(t)$ and $d\Lambda_2^{a_2}(t)$. However, $Y_3(t;r,a_3)$ is zero almost everywhere unless there are observations at $R^{a_3}=r$. To yield well-defined estimators for $\Lambda_3^{a_3}(t;s)$, processes $N_3(t;r,a_3)$ and $Y_3(t;r,a_3)$ should be refined so that $Y_3(t;r,a_3)$ is nonzero and
\begin{align*}
M_3(t;r,a_3) = \int_r^t \left\{dN_3(s;r,a_3)-Y_3(s;r,a_3)d\Lambda_3^{a_3}(s;r)\right\}
\end{align*}
is a martingale with respect to some filter. To ensure that $Y_3(t;r,a_3)$ is left-continuous, we assume that the intermediate event happens just before the primary event if $R_i=T_i$.

With all hazards $d\Lambda_1^{a_1}(t)$, $d\Lambda_2^{a_2}(t)$ and $d\Lambda_3^{a_3}(t;r)$ being well-defined and identifiable, we can estimate these quantities by martingale theory. In particular, weighted Nelson--Aalen estimators \cite{winnett2002adjusted, guo2005weighted} for cumulative hazards are given by
\begin{align*}
\widehat\Lambda_1^{a_1}(t) = \int_0^t \frac{dN_1(s;a_1)}{Y_1(s;a_1)}, \
\widehat\Lambda_2^{a_2}(t) = \int_0^t \frac{dN_2(s;a_2)}{Y_2(s;a_2)}, \
\widehat\Lambda_3^{a_3}(t;r) = \int_r^t \frac{dN_3(s;r,a_3)}{Y_3(s;r,a_3)}.
\end{align*}
Then, cumulative incidence functions can be estimated by plugging these hazards estimators into the corresponding cumulative incidence functions. Finally, we establish theoretical guarantees of these estimators as in Theorem 2, whose proof is provided in Supplementary Material B.2.

\begin{theorem} \label{thm2}
Under Assumptions \ref{consis}--\ref{dis}, if $M_3(t;s,a_3)$ is a martingale with $E\{Y_3(t;s,a_3)/n\}>0$ for $t\in[0,\tau]$, then $n^{1/2}\{\widehat{F}^{a}(\cdot)-F^{a}(\cdot)\}$ converges to $n^{1/2} \{G_1^a(\cdot) + G_2^a(\cdot) + G_3^a(\cdot)\}$, whose limiting distribution is a Gaussian process, where
\begin{align*}
G_1^a(t) &= \int_0^t \bigg[\exp\{-\Lambda_1^{a_1}(t)-\Lambda_2^{a_2}(t)\} \\
&\qquad\qquad + \int_s^t\exp\{-\Lambda_1^{a_1}(u)-\Lambda_2^{a_2}(u)-\Lambda_3^{a_3}(t;u)\}d\Lambda_2^{a_2}(u)\bigg] \frac{dM_1(s;a_1)}{Y_1(s;a_1)}, \\
G_2^a(t) &= \int_0^t \bigg[\exp\{-\Lambda_1^{a_1}(t)-\Lambda_2^{a_2}(t)\} - \exp\{-\Lambda_1^{a_1}(s)-\Lambda_2^{a_2}(s)-\Lambda_3^{a_3}(t;s)\} \\
&\qquad\qquad + \int_s^t\exp\{-\Lambda_1^{a_1}(u)-\Lambda_2^{a_2}(u)-\Lambda_3^{a_3}(t;u)\}d\Lambda_2^{a_2}(u)\bigg] \frac{dM_2(s;a_2)}{Y_2(s;a_2)}, \\
G_3^a(t) &= -\int_0^t \exp\{-\Lambda_1^{a_1}(s)-\Lambda_2^{a_2}(s)-\Lambda_3^{a_3}(t;s)\} \int_s^t\frac{dM_3(u;s,a_3)}{Y_3(u;s,a_3)}d\Lambda_2^{a_2}(s).
\end{align*}
\end{theorem}

To ensure that $M_3(t;r,a_3)$ is a well-defined martingale, we need to impose some restrictions on $d\Lambda_3^{a_3}(t;r)$. Two common choices are the Markovness $d\Lambda_3^{a_3}(t;r)=d\Lambda_{3,\text{ma.}}^{a_3}(t)$, which states that the transition rate from State 2 to State 3 depends only on the duration after State 0, and the semi-Markovness $d\Lambda_3^{a_3}(t;r)=d\Lambda_{3,\text{sm.}}^{a_3}(t-r)$, which states that the transition rate from State 2 to State 3 depends only on the duration after State 2. The choice between these two assumptions should be examined in a case-by-case manner. In general, one can always first perform a hypothesis test for the Markov assumption as described in Huang \cite{huang2021causal} to see whether a Markov assumption should be rejected or not. Moreover, as a rule of thumb, the semi-Markov assumption is often more reasonable if the intermediate event results in severe risk of terminal events. Based on these assumptions, we have the following results in Lemma \ref{lem1}.

\begin{lemma} \label{lem1}
Let $N_{3,k}(t)$, $Y_{3,k}(t)$, $Y_{3,k}^w(t)$ and $M_{3,k}(t)$ be the counting process, at-risk process, weighted at-risk process and residual of the indirect outcome event ($k = \text{ma. and sm.}$), whose expressions are given in Supplementary Material B.1.
Then
\[
d\Lambda^{a_3}_{3,k}(t) = E\left\{\int_0^t \frac{dN_{3,k}(s;a_3)}{Y_{3,k}(s;a_3)}\right\}, \quad k = \text{ma. and sm.}.
\]
The residuals $\{M_j(t;a_j): j=1,2,3 (\text{ma. and sm.})\}$ are martingales with respect to filters
\begin{align*}
\mathcal{F}_j^{a_j}(t) &= \{w_i(a_j), I(T_i^{a_j}\ge s, R_i^{a_j}\ge s, C_i^{a_j}\ge s): s \le t, i=1,\ldots,n\}, \  j=1,2, \\
\mathcal{F}_{3,\text{ma.}}^{a_3}(t) &= \{w_i(a_3), I(T_i^{a_3}\ge s, T_i^{a_3}\ge R_i^{a_3}, C_i^{a_3}\ge s): s \le t, i=1,\ldots,n\}, \\
\mathcal{F}_{3,\text{sm.}}^{a_3}(t) &= \{w_i(a_3), I(T_i^{a_3}-R_i^{a_3}\ge s, C_i^{a_3}-R_i^{a_3}\ge s): s \le t, i=1,\ldots,n\},
\end{align*}
with
$E\{dM_j(t;a_j) \mid \mathcal{F}_j^{a_j}(t)\} = 0$ and
$\var\{dM_j(t;a_j) \mid \mathcal{F}_j^{a_j}(t)\} = Y_j^w(t;a_j)d\Lambda_j^{a_j}(t)$.
\end{lemma}

Lemma \ref{lem1} illustrates that $d\Lambda_3^a(t;r)$ can be estimated under either Markovness or semi-Markovness. Its proof is provided in Supplementary Material B.1. When there is no confusion, we omit the subscript ``ma." or ``sm." in $N_3(\cdot;a_3)$, $Y_3(\cdot;a_3)$, $M_3(\cdot;a_3)$ and $d\Lambda_3^{a_3}(t)$. 
As a regularity condition on positivity, assume $Y^{-1}_j(\tau;a_j) = o(n^{-1/2})$ for $j = 1,2,3 (\text{ma. and sm.})$ in the following.

\begin{corollary}
Under Assumptions \ref{consis}--\ref{dis}, if the hazard $d\Lambda_3^{a_3}(t;r)$ is Markov, i.e., $d\Lambda_3^{a_3}(t;r) = d\Lambda_{3}^{a_3}(t)$ for every $r\in[0,\tau]$, then $n^{1/2}\{\widehat{F}^{a}(t)-F^{a}(t)\}\stackrel{d}{\longrightarrow}N\{0,\sigma^2(t)\}$ on $t\in[0,\tau]$, with
\begin{align*}
\sigma^2(t) &= \int_0^t \left[1-F^a(t)-\{F_2^a(s)-F_3^a(s)\}\exp\{\Lambda_3^{a_3}(s)-\Lambda_3^{a_3}(t)\}\right]^2 \frac{E\{Y_1^w(s;a_1)/n\}}{[E\{Y_1(s;a_1)/n\}]^2}d\Lambda_1^{a_1}(s) \\
&\quad+ \int_0^t \left[1-F^a(t)-\{1-F^a(s)\}\exp\{\Lambda_3^{a_3}(s)-\Lambda_3^{a_3}(t)\}\right]^2 \frac{E\{Y_2^w(s;a_2)/n\}}{[E\{Y_2(s;a_2)/n\}]^2}d\Lambda_2^{a_2}(s) \\
&\quad+ \int_0^t \left[\{F_2^a(s)-F_3^a(s)\}\exp\{\Lambda_3^{a_3}(s)-\Lambda_3^{a_3}(t)\}\right]^2 \frac{E\{Y_3^w(s;a_3)/n\}}{[E\{Y_3(s;a_3)/n\}]^2}d\Lambda_{3}^{a_3}(s),
\end{align*}
where $\stackrel{d}{\longrightarrow}$ denotes converging in distribution.
\end{corollary}

\begin{corollary}
Under Assumptions \ref{consis}--\ref{dis}, if the hazard $d\Lambda_3^{a_3}(t;r)$ is semi-Markov, i.e., $d\Lambda_3^{a_3}(t;r) = d\Lambda_{3}^{a_3}(t-r)$, then $n^{1/2}\{\widehat{F}^{a}(t)-F^{a}(t)\}\stackrel{d}{\longrightarrow}N\{0,\sigma^2(t)\}$ on $t\in[0,\tau]$, with
\begin{align*}
\sigma^2(t) &= \int_0^t \left[1-F_1^a(t)-F_2^a(t)+\int_s^t\exp\{-\Lambda_3^{a_3}(t-u)\}dF_2^a(u)\right]^2 \frac{E\{Y_1^w(s;a_1)/n\}}{[E\{Y_1(s;a_1)/n\}]^2}d\Lambda_1^{a_1}(s) \\
&\quad + \int_0^t \bigg[\{1-F_1^a(u)-F_2^a(u)\}\exp\{-\Lambda_3^{a_3}(t-u)\}\Big{|}_{s}^{t} \\
&\qquad\qquad\quad + \int_s^t\exp\{-\Lambda_3^{a_3}(t-u)\}dF_2^a(u)\bigg]^2 \frac{E\{Y_2^w(s;a_2)/n\}}{[E\{Y_2(s;a_2)/n\}]^2}d\Lambda_2^{a_2}(s) \\
&\quad + \int_0^t \left[\int_0^{t-s} \exp\{-\Lambda_3^{a_3}(t-u)\}dF_2^a(u)\right]^2 \frac{E\{Y_3^w(s;a_3)/n\}}{[E\{Y_3(s;a_3)/n\}]^2}d\Lambda_{3}^{a_3}(s).
\end{align*}
\end{corollary}

A consistent estimator for $\sigma^2(t)$ can be obtained by plugging the estimated hazards in the expression of $\sigma^2(t)$. Hypothesis tests for separable effects can be performed using logrank statistics. Test statistics with asymptotics are provided in Supplementary Material D.

When the true $d\Lambda_3^{a_3}(t;r)$ satisfies neither Markovness or semi-Markovness, a sensitivity analysis is proposed by assuming a mixture of Markovness and semi-Markovness. The following theorem gives the uniform convergence of the estimated counterfactual cumulative incidence under additional assumptions, whose proof is provided in Supplementary Material C. If the mixture pattern is unknown, likelihood methods may be adopted to estimate the mixed hazards.

\begin{theorem} \label{thm3}
Suppose $d\Lambda_3^{a_3}(t;r)$ is a linear combination of $d\Lambda_{3,\text{ma.}}^{a_3}(t)$ and $d\Lambda_{3,\text{sm.}}^{a_3}(t-r)$,
\begin{align}
d\Lambda_3^{a_3}(t;r) = (1-\kappa)d\Lambda_{3,\text{ma.}}^{a_3}(t) + \kappa d\Lambda_{3,\text{sm.}}^{a_3}(t-r)
\end{align}
where $\kappa\in[0,1]$ is a prespecified parameter. Under Assumptions \ref{consis}--\ref{dis},
$\sup_{t \in [0,\tau]} |\widehat{F}^{a}(t)-F^{a}(t) | \stackrel{p}{\longrightarrow} 0$ if $\kappa$ is correctly specified.
\end{theorem}

\section{Estimation under the general case} \label{sec4}

It is more challenging to nonparametrically estimate the counterfactual cumulative incidences if covariates isolation fails. The first reason is that we need to employ models for the cause-specific hazards \cite{breum2024estimation}. The second reason is that there is a biased sampling issue when Markovness does not hold. Given the history at a time, the potential censoring time is not independent of the potential time to the indirect outcome event anymore.

Let $\mathcal{X}$ be the support of $X$. It is obvious that $d\Lambda_1^{a_1}(t;x)$ and $d\Lambda_2^{a_2}(t;x)$ are estimable, for example, using proportional hazards or additive hazards models. In order that $d\Lambda_3^{a_3}(t;r,x)$ is estimable, we assume that there is a subset of history $\mathcal{H}(t;r)$ such that $d\Lambda_3^{a_3}(t;r,x)$ is identical for $(t,r)\in\mathcal{H}(t,r)$ given $X=x$. Markvoness (conditional on covariates) implies $\mathcal{H}(t;r) = \{(\widetilde{T},\widetilde{R}): \widetilde{T}\geq t, \widetilde{R}\leq t\}$, and semi-Markovness (conditional on covariates) implies $\mathcal{H}(t;r) = \{(\widetilde{T},\widetilde{R}): \widetilde{T}-\widetilde{R}\geq t-r\}$. We modify the martingales by multiplying the indicator $I(X=x)$,
\[
M_1(t;a_1,x), \ M_2(t;a_2,x), \ M_3(t;r,a_3,x).
\]
The at-risk set in the martingale $M_3(t;r,a_3,x)$ conditional on $X=x$ has probability $P((\widetilde{T},\widetilde{R})\in\mathcal{H}(t;r), A=a_3 \mid X=x)$.

\begin{theorem} \label{thm4}
The efficient influence function (EIF) of $F^{(a_1,a_2,a_3)}(t)$ is
\begin{align*}
\varphi^{(a_1,a_2,a_3)}(t) &= \int_0^t \exp\{-\Lambda_1^{a_1}(s;X)\} \bigg\{\frac{I(A=a_1)}{P(A=a_1\mid X)} \frac{dM_1(s;A,X)}{P(\widetilde{T}\wedge\widetilde{R}\geq s \mid A,X)} \\
&\quad - \sum_{j\in\{1,2\}}\frac{I(A=a_j)}{P(A=a_j\mid X)} \int_0^s \frac{dM_j(u;A,X)}{P(\widetilde{T}\wedge\widetilde{R}\geq u\mid A,X)} d\Lambda_1^{a_1}(s;X)\bigg\} \\
&\quad + \int_0^t\int_0^s \exp\{-\Lambda_1^{a_1}(r;X)-\Lambda_2^{a_2}(r;X)-\Lambda_3^{a_3}(s;r,X)\} \\
&\quad \bigg\{\frac{I(A=a_2)}{P(A=a_2\mid X)} \frac{dM_2(r;A,X)}{P(\widetilde{T}\wedge\widetilde{R}\geq r)} d\Lambda_3^{a_3}(s;r,X) \\
&\quad - \sum_{j\in\{1,2\}} \frac{I(A=a_j)}{P(A=a_j\mid X)} \int_0^r \frac{dM_j(u;A,X)}{P(\widetilde{T}\wedge\widetilde{R}\geq u \mid A,X)} d\Lambda_2^{a_2}(r;X)d\Lambda_3^{a_3}(s;r,X) \\
&\quad + \frac{I(A=a_3)}{P(A=a_3\mid X)} \frac{dM_3(s;\widetilde{R},A,X)}{P(\mathcal{H}(s,\widetilde{R})\mid A,X)} d\Lambda_2^{a_2}(s;X) \\
&\quad - \frac{I(A=a_3)}{P(A=a_3\mid X)} \int_{\mathcal{H}(s,\widetilde{R})}\frac{dM_3(u;\widetilde{R},A,X)}{P(\mathcal{H}(u,\widetilde{R})\mid A,X)} d\Lambda_2^{a_2}(r;X)d\Lambda_3^{a_3}(s;r,X) \bigg\} \\
&\quad + F^{(a_1,a_2,a_3)}(t;X) - F^{(a_1,a_2,a_3)}(t).
\end{align*}
\end{theorem}

We impose semiparametric models for the propensity score, casue-specific hazards and censoring hazard. By plugging in the fitted models in the efficient influence function and solving the equation that the empirical mean $P_n\{\widehat\varphi^a(t)\}=0$, we obtain the estimate of $F^{(a_1,a_2,a_3)}(t)$, denoted by $\widetilde{F}^a(t)$. The resulting estimator has multiple robustness in that it is consistent if (1) all the three cause-specific hazards are correctly specified, or (2) the propensity score the censoring hazard are correctly specified and at most one cause-specific hazard is misspecified \cite{breum2024estimation}. Specially, under (conditional) Markovness,
\begin{align*}
\widetilde{F}^{(a_1,a_2,a_3)}(t) &= P_n \bigg[\int_0^t \exp\{-\widehat\Lambda_1^{a_1}(s;X)\} \bigg\{\frac{I(A=a_1)}{\widehat{P}(A=a_1\mid X)} \frac{d\widehat{M}_1(s;A,X)}{\exp\{-\widehat\Lambda_1^A(s;X)-\widehat\Lambda_1^A(s;X)-\widehat\Lambda_C^A(s;X)\}} \\
&\qquad - \sum_{j\in\{1,2\}}\frac{I(A=a_j)}{\widehat{P}(A=a_j\mid X)} \int_0^s \frac{d\widehat{M}_j(u;A,X) d\widehat\Lambda_1^{a_1}(s;X)}{\exp\{-\widehat\Lambda_1^A(u;X)-\widehat\Lambda_2^A(u;X)-\widehat\Lambda_C^A(u;X)\}} \bigg\} \\
&\qquad + \int_0^t\int_0^s \exp\{-\widehat\Lambda_1^{a_1}(r;X)-\widehat\Lambda_2^{a_2}(r;X)-\widehat\Lambda_3^{a_3}(s;X)+\widehat\Lambda_3^{a_3}(r;X)\} \\
&\qquad \bigg\{\frac{I(A=a_2)}{\widehat{P}(A=a_2\mid X)} \frac{d\widehat{M}_2(r;A,X) d\widehat\Lambda_3^{a_3}(s;X)}{\exp\{-\widehat\Lambda_1^A(r;X)-\widehat\Lambda_2^A(r;X)-\widehat\Lambda_C^A(r;X)\}} \\
&\qquad - \sum_{j\in\{1,2\}} \frac{I(A=a_j)}{\widehat{P}(A=a_j\mid X)} \int_0^r \frac{d\widehat{M}_j(u;A,X) d\widehat\Lambda_2^{a_2}(r;X)d\widehat\Lambda_3^{a_3}(s;X)}{\exp\{-\widehat\Lambda_1^A(u;X)-\widehat\Lambda_2^A(u;X)-\widehat\Lambda_C^A(u;X)\}}  \\
&\qquad + \frac{I(A=a_3)}{\widehat{P}(A=a_3\mid X)} \frac{d\widehat{M}_3(s;\widetilde{R},A,X) d\widehat\Lambda_2^{a_2}(s;X)}{\int_0^s\exp\{-\widehat\Lambda_1^{A}(u;x)-\widehat\Lambda_2^{A}(u;x)-\widehat\Lambda_3^A(s;x)-\widehat\Lambda_C^A(s;x)+\widehat\Lambda_3^A(u;x)\}d\widehat\Lambda_2^A(u;x)} \\
&\qquad - \frac{I(A=a_3)}{\widehat{P}(A=a_3\mid X)} \int_{\widetilde{R}}^s\frac{d\widehat{M}_3(u;\widetilde{R},A,X) d\widehat\Lambda_2^{a_2}(r;X)d\widehat\Lambda_3^{a_3}(s;X)}{\int_0^u\exp\{-\widehat\Lambda_1^{A}(v;x)-\widehat\Lambda_2^{A}(v;x)-\widehat\Lambda_3^A(v;x)-\widehat\Lambda_C^A(u;x)+\widehat\Lambda_3^A(u;x)\}d\widehat\Lambda_2^A(v;x)} \bigg\} \\
&\qquad + \int_0^t \exp\{-\widehat\Lambda_1^{a_1}(s;X)-\widehat\Lambda_2^{a_2}(s;X)\}d\widehat\Lambda_1^{a_1}(s;X) \\
&\qquad + \int_0^t\int_0^s \exp\{-\widehat\Lambda_1^{a_1}(r;X)-\widehat\Lambda_2^{a_2}(r;X)-\widehat\Lambda_3^{a_3}(s;X)+\widehat\Lambda_3^{a_3}(r;X)\}d\widehat\Lambda_2^{a_2}(r;X)d\widehat\Lambda_3^{a_3}(s;X) \bigg],
\end{align*}
where $\widehat\Lambda_C^A(t;x)$ is the fitted cumulative hazard of censoring.
When all models are correctly specified, the EIF-based estimator has asymptotic normality.

\begin{theorem} \label{thm5}
Assume Assumptions \ref{consis}, \ref{ign}, \ref{ran_pi}, \ref{pos_pi} and \ref{dis_pi} hold. In addition, assume the fitted model for $\varphi^a(t)$ belongs to a Donsker class, and all the fitted submodels (propensity score, cause-specific hazards and censoring hazard) converge at rate faster than $o_p(n^{-1/2})$, then $n^{1/2}\{\widetilde{F}^{a}(t)-F^{a}(t)\}\stackrel{d}{\longrightarrow}N\{0,\sigma^2(t)\}$ on $t\in[0,\tau]$, with $\sigma^2(t) = E[\{\varphi^a(t)\}^2]$.
\end{theorem}

The proof is given in Supplementary Material E. Even if the models are fitted with uncertainty, the asymptotic property is still valid. However, estimation is computational challenging especially if we do not assume Markovness. If we do not know whether the hazard of the indirect outcome event is Markov, but as long as all other models (propensity score, cause-specific hazards of direct outcome event and intermediate event, censoring hazard) are correctly specified, the EIF-based estimator will be consistent. The asymptotic variance may be incorrect.

 Hypothetis test for the separable pathway effect of $(a_1,a_2,a_3)$ against $(a_1',a_2',a_3')$ can be constructed based on the statistic
\begin{equation}
U = \int_0^{\tau} \{\widetilde{F}^{a}(t)-\widetilde{F}^{a'}(t)\} d\{\widetilde{F}^{a}(t)+\widetilde{F}^{a'}(t)\}. \label{testsu}
\end{equation}
The asymptotic variance of $U$ can be calculated by the functional delta method based on the influence function. In Supplementary Material G, we derive the identifiability of the counterfactual cumulative incidence in the presence of post-treatment time-varying covariates.

\section{Simulation studies} \label{sec6}

In this section, we conduct simulation studies to demonstrate the proposed methods in estimating counterfactual cumulative incidences and testing separable pathway effects. We assume an additive hazards model so that the generalized Nelson--Aalen estimator by inverse probaility weighting (GNAIPW) in Section \ref{sec3} is valid.

We generate a sample with $n = 500$ independent units. For $i\in\{1,\ldots,n\}$, we generate two dichotomized covariates $X_i = (X_{i1}, X_{i2})$, each equals 1 or 0.5 with equal probability. A unit receives treatment with probability $P(A_i=1 \mid X_i) = \expit(0.4X_{i1}+0.8X_{i2}-0.6)$, where $\expit(x)=1/\{1+\exp(-x)\}$. Consider three settings of hazards ($a$ is a scalar):
\begin{enumerate}
\item Setting 1: $d\Lambda_1^{a}(t;x) = 0.15(x_1+a)dt$, $d\Lambda_2^{a}(t;x) = 0.1(x_1+a)dt$, $d\Lambda_3^{a}(t;x) = 0.2(x_2+a)dt$. Both the Markov assumption and semi-Markov assumption are satisfied.
\item Setting 2: $d\Lambda_1^{a}(t;x) = 0.04(x_1+a)tdt$, $d\Lambda_2^{a}(t;x) = 0.02(x_1+a)tdt$, $d\Lambda_3^{a}(t;x) = 0.05(x_2+a)tdt$. Only the Markov assumption is satisfied.
\item Setting 3: $d\Lambda_1^{a}(t;x) = 0.04(x_1+a)tdt$, $d\Lambda_2^{a}(t;x) = 0.02(x_1+a)tdt$, $d\Lambda_3^{a}(t;r,x) = 0.1(x_2+a)(t-r)dt$. Only the semi-Markov assumption is satisfied.
\end{enumerate}
Assumptions \ref{consis}--\ref{dis} hold under all these three settings. We evaluate the estimation performance of our method for making inference on the following counterfactual cumulative incidences: $F^{(1,0,0)}(t)$ and $F^{(1,0,1)}(t)$. These two functions measure the counterfactual cumulative incidence by intervening the hazard directly to primary event and by intervening the hazards both directly and indirectly to the primary event, respectively. These two quantities are of our interest because different versions of direct treatment effects can be obtained by contrasting these two quantities with the cumulative incidence in the real world.

\begin{figure}
\centering
\includegraphics[width=0.95\textwidth]{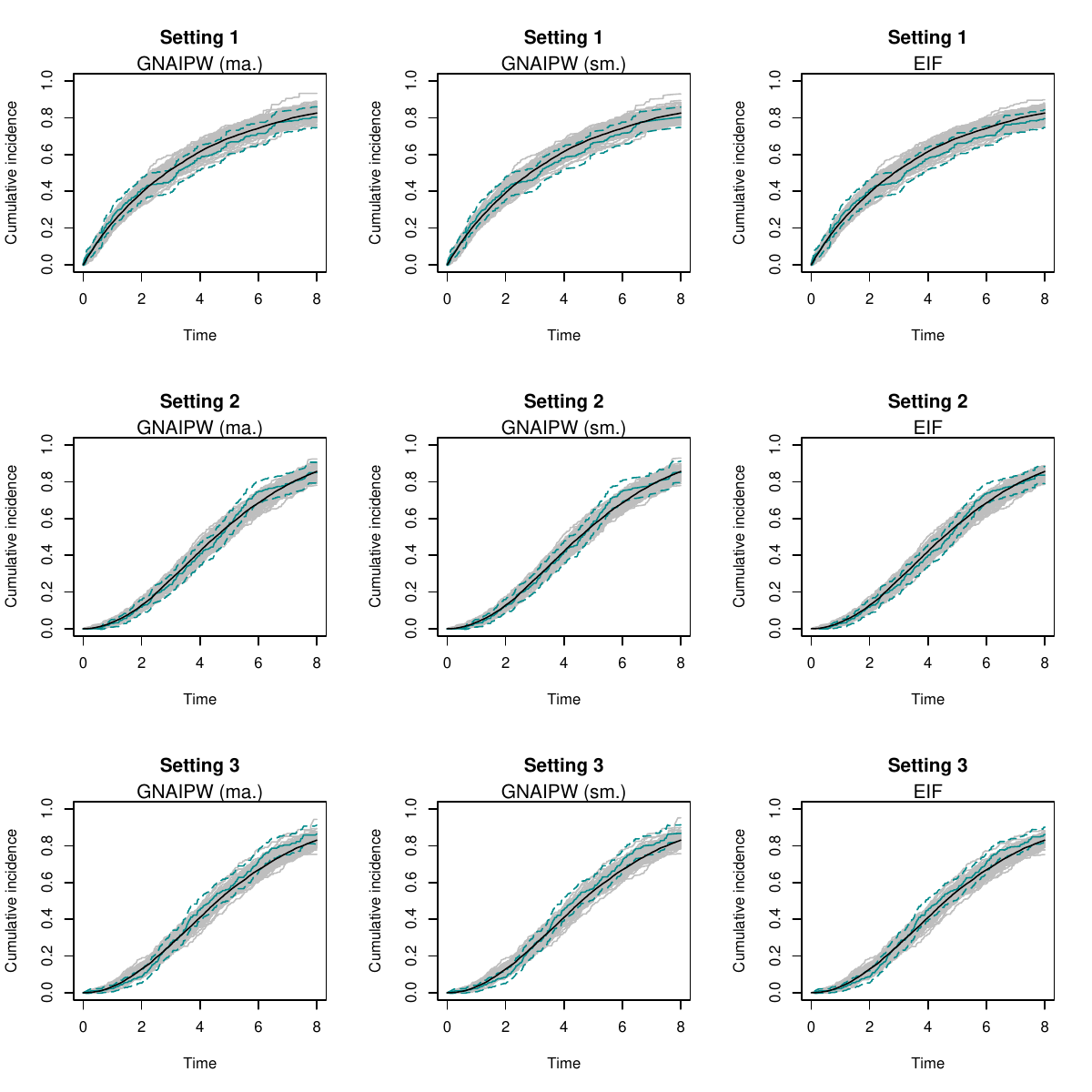}
\caption{Estimated cumulative incidence functions for $F^{(1,0,0)}(t)$. In each panel, the black line is the true incidence, each grey line is an estimated incidence, the solid cyan line is a randomly chosen estimated incidence, and dashed cyan lines denote the 95\% asymptotic confidence interval.}
\label{curve1}
\end{figure}

\begin{figure}
\centering
\includegraphics[width=0.95\textwidth]{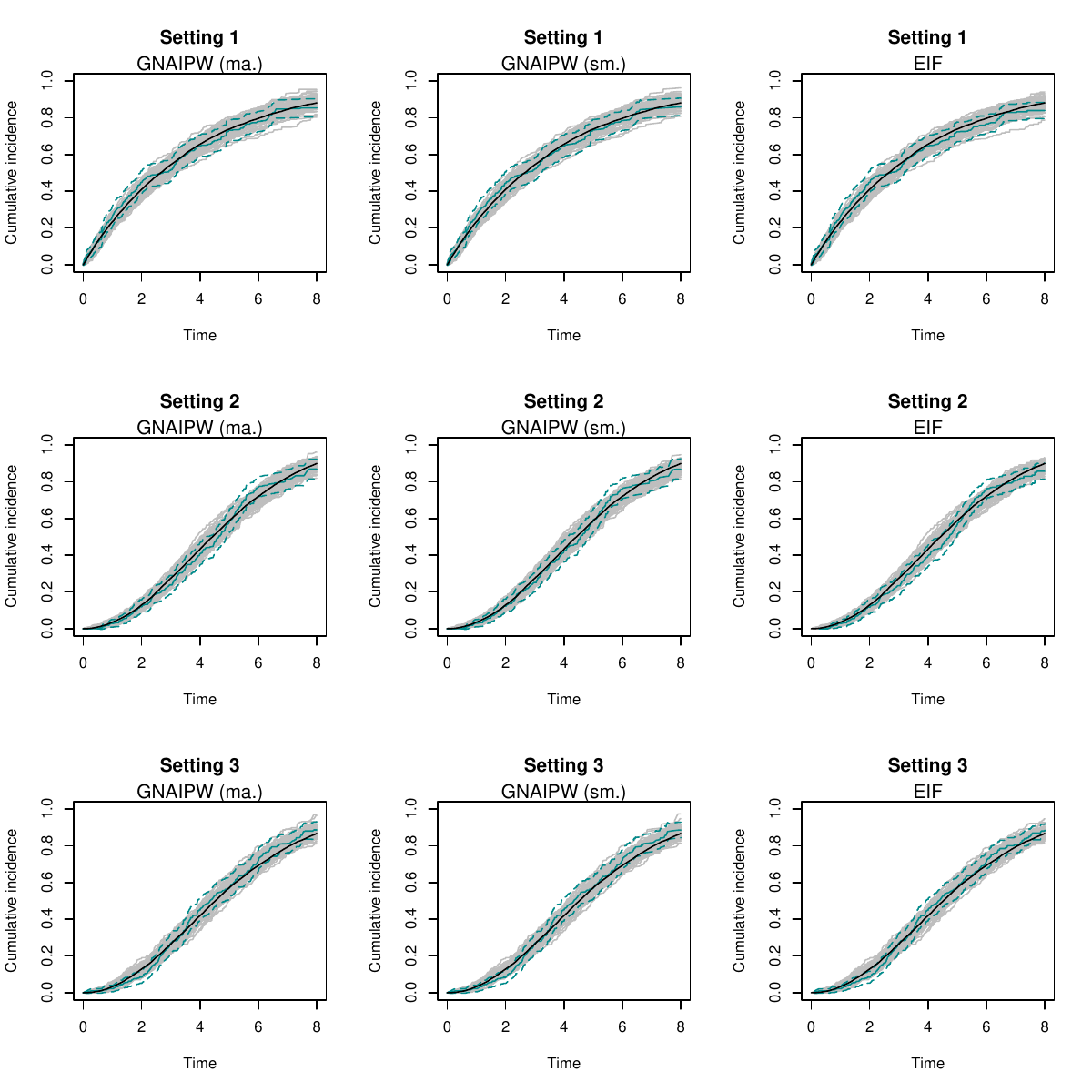}
\caption{Estimated cumulative incidence functions for $F^{(1,0,1)}(t)$. In each panel, the black line is the true incidence, each grey line is an estimated incidence, the solid cyan line is a randomly chosen estimated incidence, and dashed cyan lines denote the 95\% asymptotic confidence interval.}
\label{curve2}
\end{figure}

Figures \ref{curve1}--\ref{curve2} show the estimated counterfactual cumulative incidences for $F^{(1,0,0)}(t)$ and $F^{(1,0,1)}(t)$, respectively. In each panel, the black line represents the true incidence $F^{(a_1,a_2,a_3)}(t)$. To see the variation of estimators and the influence of adopting Markovness or semi-Markovness in estimation, estimated cumulative incidence curves based on 100 independently generated datasets are drawn in grey color. The first column displays estimated curves using GNAIPW estimator with Markovness, and the second column using  GNAIPW estimator with semi-Markovness. The third column displays estimated curves using the EIF-based estimator. In each sub-figure, we randomly choose an estimated curve in cyan color. The dashed cyan line shows the pointwise 95\% confidence interval by using the asymptotic formula. The confidence intervals generally cover the true line.


Additional simulation results including estimates and biases are shown in Supplementary Material H.1. With the sample size being larger, the estimated curves show smaller variations. The proposed methods (GNAIPW and EIF) are compared with the existing method of Huang \cite{huang2021causal}. Supplementary Material H.2 provides assessments of confidence intervals by the asymptotic formula. When the sample size is moderate or large, the coverage rate of the proposed method is close to the nominal level. Supplementary Material H.3 assesses the performance of hypothesis tests. Supplementary Material H.4 conducts simulation studies when covariates isolation fails but the conditional dismissible components assumption holds. Supplementary Material H.5 conducts a sensitivity analysis when the treatment components have interaction.

\section{Application to Allogeneic Stem Cell Transplantation Data} \label{sec7}

Allogeneic stem cell transplantation is a well applied therapy to treat acute lymphoblastic leukemia (ALL), including two sorts of transplant modalities: human leukocyte antigens matched sibling donor transplantation (MSDT) and haploidentical stem cell transplantation from family (Haplo-SCT). MSDT has long been regarded as the first choice of transplantation because MSDT leads to lower transplant-related mortality, also known as non-relapse mortality (NRM) \cite{kanakry2016modern}. Another source of mortality is due to relapse, known as relapse related mortality (RRM). In recent years, some benefits of Haplo-SCT have been noticed that patients with positive pre-transplantation minimum residual disease (MRD) undergoing Haplo-SCT have better prognosis in relapse \cite{chang2020haploidentical}. The contradictory effects of transplant modalities on NRM and relapse motivate us to investigate how different transplant modalities exert effects on the overall survival.


A total of $n=239$ patients with positive MRD in first complete remission (CR1) undergoing allogeneic stem cell transplantation are included in our study \cite{ma2021an}. Among these patients, 55 received MSDT ($A_i=0$) and 184 received Haplo-SCT ($A_i=1$). The transplantation type is ``genetically randomized'' in that there is no specific consideration to prefer Haplo-SCT over MSDT whenever MSDT is accessible \cite{chang2020haploidentical}, so there should be no other covariates apart from the accessibility of MSDT that affect the treatment and outcomes simultaneously. The accessibility is related with age, as older people are more likely to have MSDT donors due to the ``one-child'' policy in China. We expect ignorability by conditioning on sex, age and diagnosis (T-ALL or B-ALL). The propensity score is fitted by logistic regression. Let $R_i$ be the time of relapse and $T_i$ be the time of death after transplantation. In the MSDT group, 43.6\% individuals were observed to experience relapse and 50.9\% mortality. In the Haplo-SCT group, 25.5\% individuals were observed to experience relapse and 31.0\% mortality. Table \ref{sumstat} lists the number of four possible state paths categorized by treatment arm.

\begin{table}
    \centering
    \caption{Observed event statuses categorized by treatment groups}
    \begin{tabular}{cccccl}
    \toprule
    $\delta^R$ & $\delta^T$ & MSDT & Haplo-SCT & Total & Description \\
    \midrule
    1 & 1 & 22 & 37 & 59 & Dead with relapse \\
    1 & 0 & 2 & 10 & 12 & Censored with relapse \\
    0 & 1 & 6 & 20 & 26 & Dead without relapse \\
    0 & 0 & 25 & 117 & 142 & Censored without relapse or death \\
    \bottomrule
    \end{tabular}
    \label{sumstat}
\end{table}

Transplant-related mortality (non-relapse mortality) is mainly caused by infection due to low immunity. Since there are mismatched HLA loci if receiving Haplo-SCT, there would be stronger acute graft-versus-host disease (GVHD) with Haplo-SCT. In practice, patients receiving Haplo-SCT should additionally use antithymocyte globulin (ATG) to facilitate engraftment \cite{walker2016pretreatment, giebel2023posttransplant}. Therefore, $A_1$ is a component through delaying immune reconstitution (the combined usage of ATG), which increases the risk of transplant-related mortality. This is also why MSDT is preferred over Haplo-SCT \cite{kanakry2016modern}. Leukemia relapse is caused by the presence of minimum residual disease (MRD). The stronger immune rejection with Haplo-SCT also kills the minimum residual disease cells, which is referred to as the ``graft-versus-leukemia'' effect \cite{chang2020haploidentical}. Therefore, $A_2$ is a component through eradicating MRD, which reduces the risk of relapse. Patients with relapse have fewer normal leukocyte cells, and relapse-related mortality is mainly caused by hemorrhagic syndrome or infection caused by the dramatic decline of normal leukocyte. Therefore, $A_3$ is a component indicating the long-term effect on mortality through the decline of leukocyte. This effect has not been studied by previous literature.

Previous multivariate analyses indicated that age is only related with RRM and diagnosis (T-ALL or B-ALL) is only related with relapse \cite{hangai2019allogeneic, chang2020haploidentical}. In fact, relapse is due to the minimum residual disease (MRD), non-relapse mortality is due to acute graft-versus-host disease and infection cause by low immunity, and relapse-related mortality is due to hemorrhagic syndrome or infection caused by the dramatic decline of normal leukocyte. No other covariates that affect NRM, relapse or RRM simultaneously were found. The effects of covariates on NRM, relapse and RRM are isolated, so we adopt the dismissible components assumption with covariates isolation. 
In the following analysis we maintain the semi-Markov assumption, because RRM usually happens soon after relapse, making it reasonable to assume that the hazard of RRM after relapse relies on how long it passed after relapse rather than the duration from transplantation to relapse.

\begin{figure}
\centering
\includegraphics[width=0.32\textwidth]{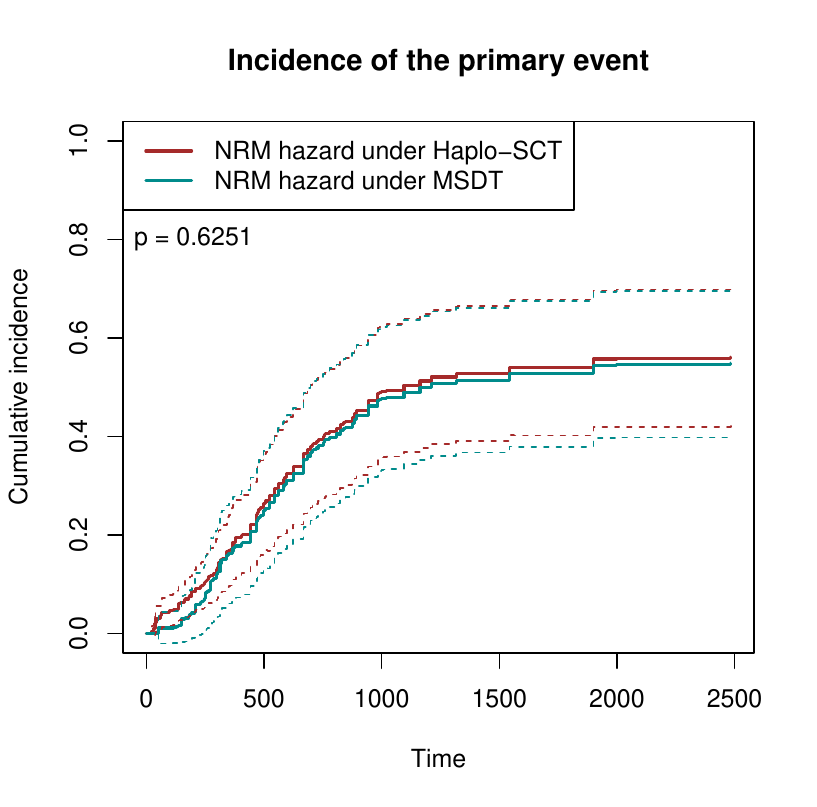}
\includegraphics[width=0.32\textwidth]{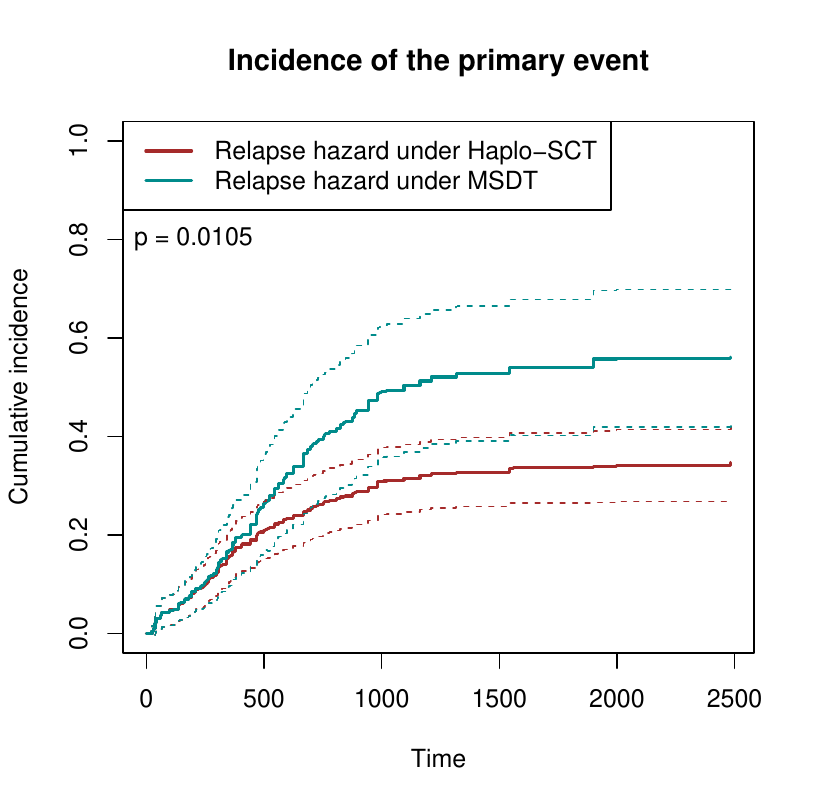}
\includegraphics[width=0.32\textwidth]{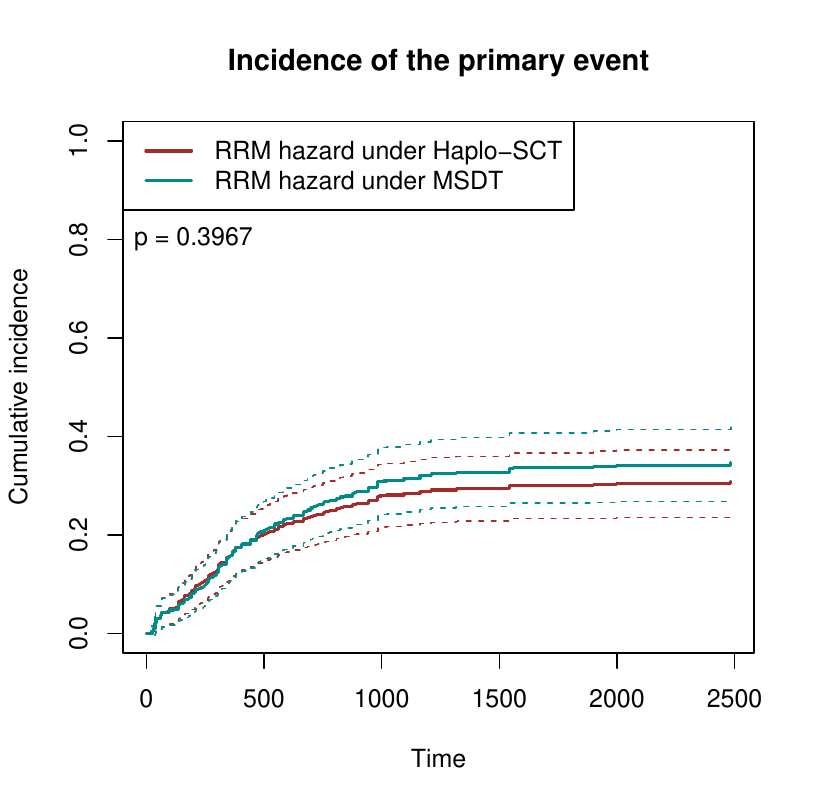} \\
\includegraphics[width=0.32\textwidth]{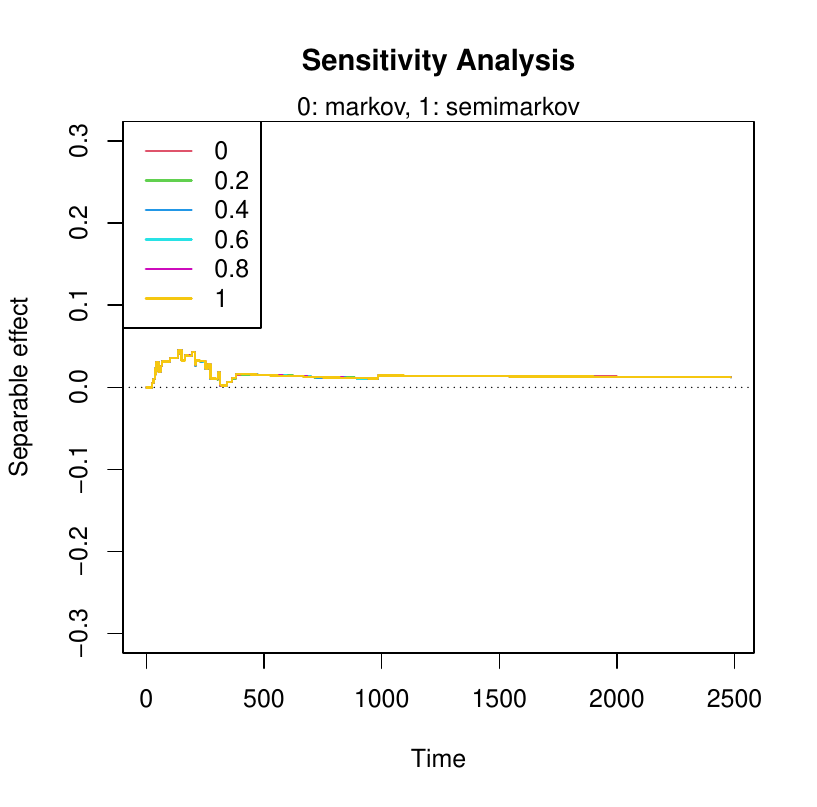}
\includegraphics[width=0.32\textwidth]{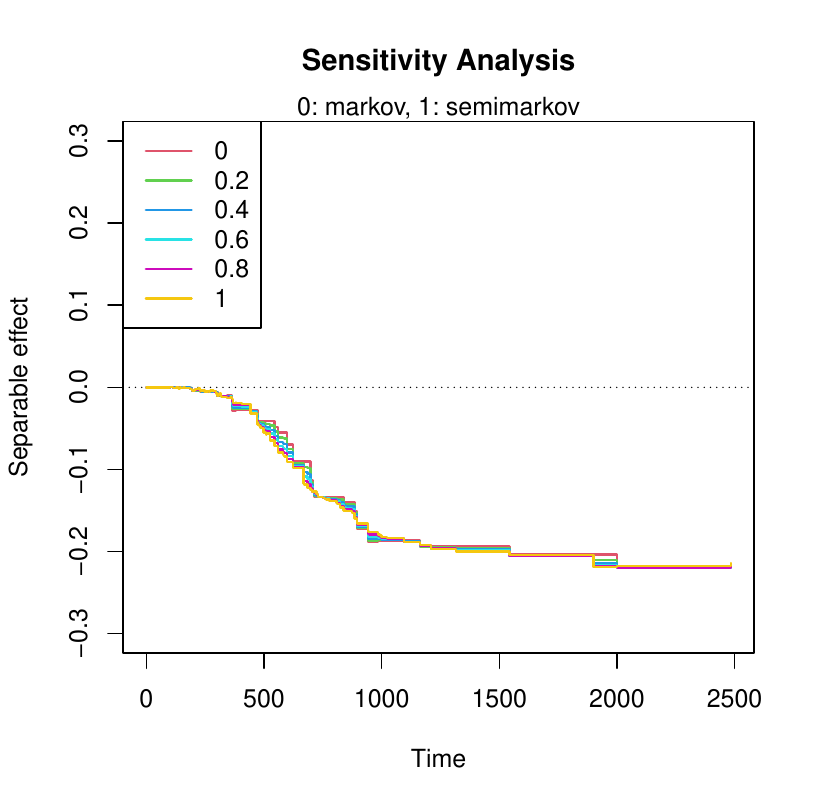}
\includegraphics[width=0.32\textwidth]{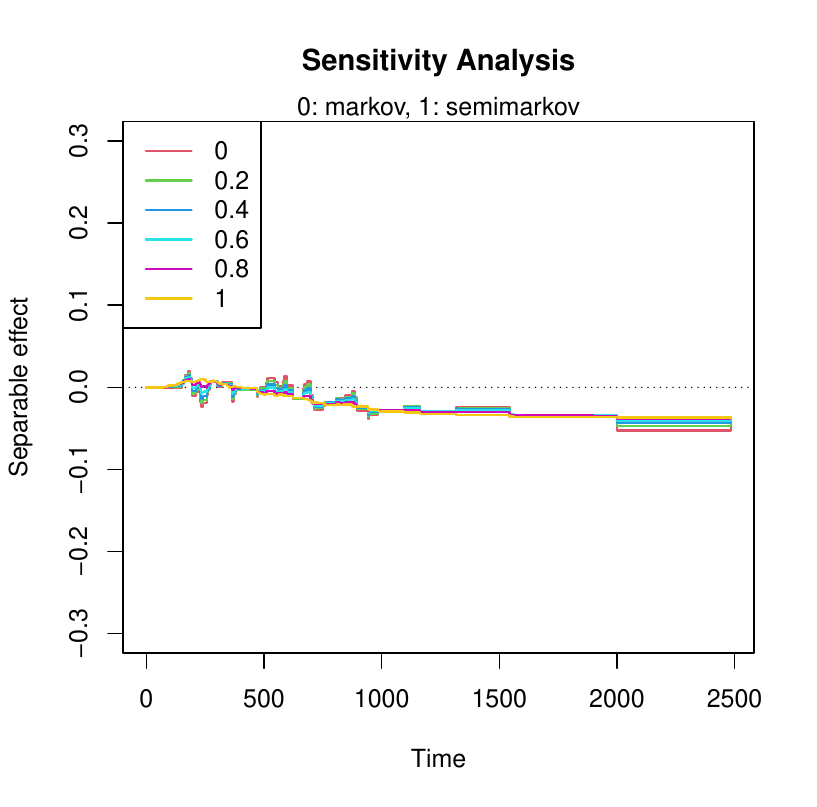} \\
\includegraphics[width=0.32\textwidth]{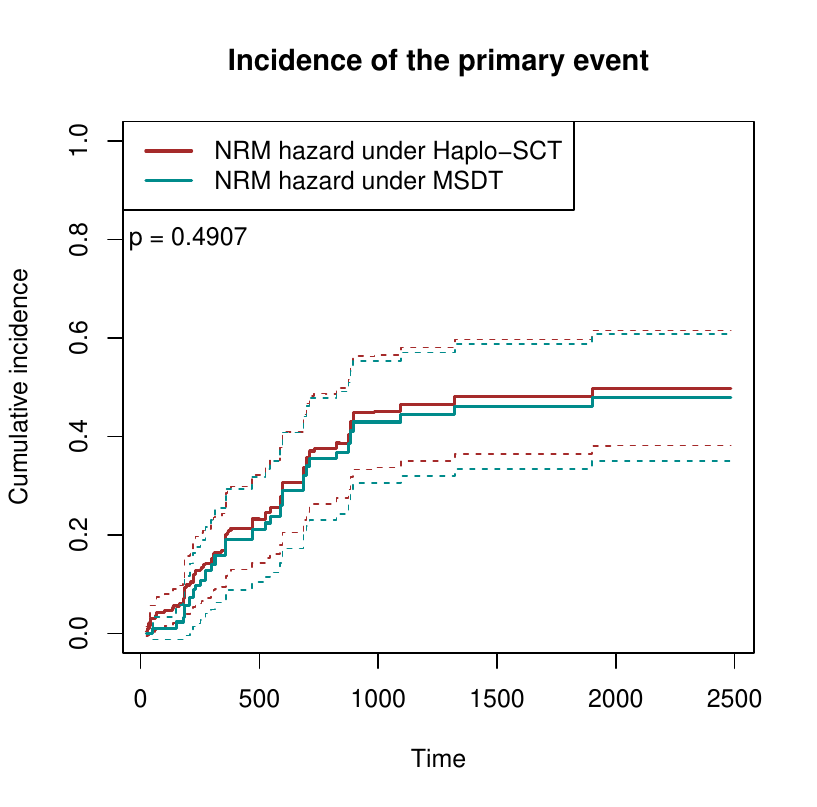}
\includegraphics[width=0.32\textwidth]{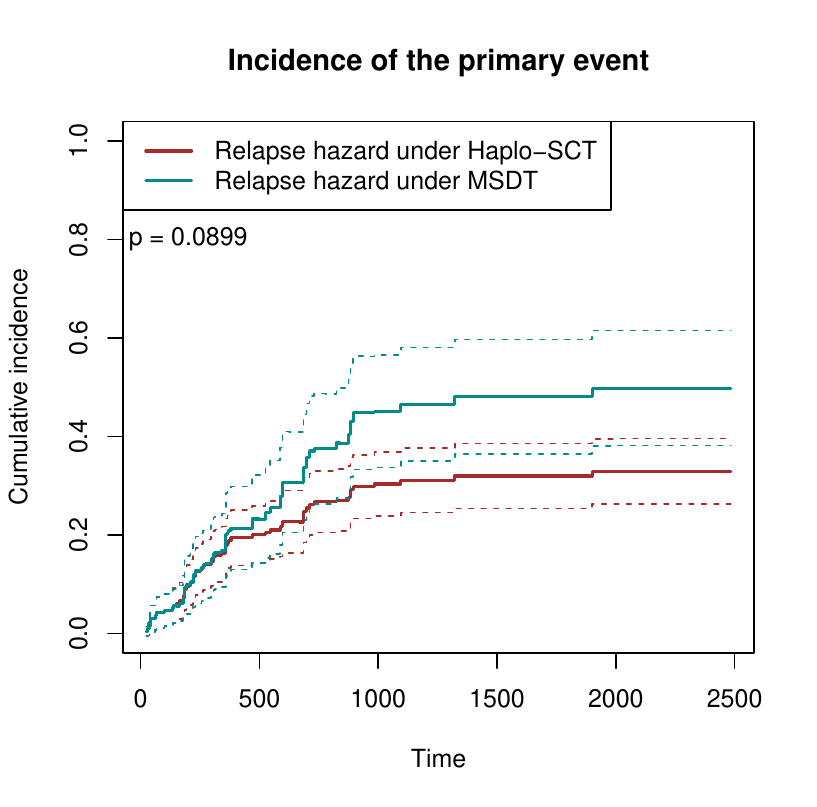}
\includegraphics[width=0.32\textwidth]{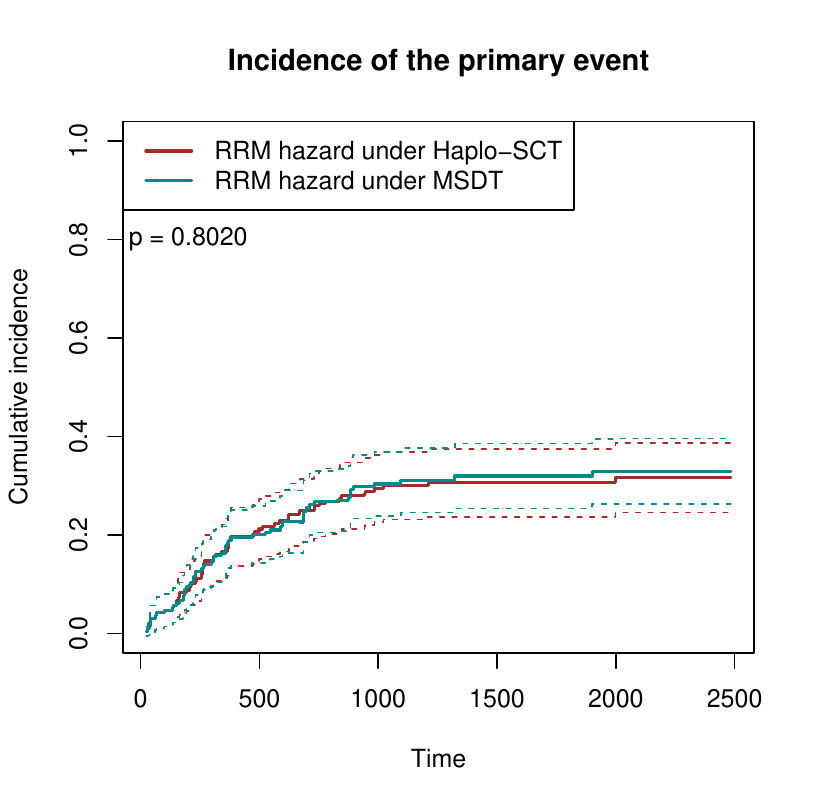}
\caption{First row: estimated counterfactual cumulative incidences of mortality (with 95\% confidence intervals) assuming covariates isolation. Second row: sensitivity analysis of separable pathway effects. Third row: estimated estimated counterfactual cumulative incidences of mortality (with 95\% confidence intervals) in the general case. Compared between (1) $F^{(1,0,0)}(t)$ and $F^{(0,0,0)}(t)$, (2) $F^{(1,1,0)}(t)$ and $F^{(1,0,0)}(t)$, (3) $F^{(1,1,1)}(t)$ and $F^{(1,1,0)}(t)$.} \label{data}
\end{figure}

In the multi-state model, States 1, 2 and 3 refer to NRM, relapse and RRM, respectively. The total effect of mortality can be decomposed into three separable pathway effects: one through NRM $\SPE_{0\to1}(t;0,0)$, one through relapse $\SPE_{0\to2}(t;1,0)$, and the other through RRM $\SPE_{2\to3}(t;1,1)$. The first row of Figure \ref{data} displays the estimated counterfactual cumulative incidence of mortality based on separable pathway effects with 95\% confidence intervals assuming covariates isolation. The first panel controls the hazards of relapse and RRM, and compares the counterfactual cumulative incidences of mortality with different hazards of NRM, i.e., $F^{(1,0,0)}(t)$ versus $F^{(0,0,0)}(t)$. The second panel controls the hazards of NRM and RRM, and compares the counterfactual cumulative incidences of mortality with different hazards of relapse, i.e., $F^{(1,1,0)}(t)$ versus $F^{(1,0,0)}(t)$. The third panel controls the hazards of NRM and relapse, and compares the counterfactual cumulative incidences of mortality with different hazards of RRM, i.e., $F^{(1,1,1)}(t)$ versus $F^{(1,1,0)}(t)$.

The estimated time-varying treatment effect by the GNAIPW estimator under covariates isolation is displayed in the second row of Figure \ref{data}. To assess the sensitivity of the assumption on $d\Lambda_3^{a_3}(t;r)$, we conduct a sensitivity analysis. Suppose
$d\Lambda_3^{a_3}(t;r) = (1-\kappa) d\Lambda_{3,\text{ma.}}^{a_3}(t) + \kappa d\Lambda_{3,\text{sm.}}^{a_3}(t-r)$.
Note that $\kappa=0$ corresponds to Markovness and $\kappa=1$ corresponds to semi-Markovness. The first separable pathway effect $\SPE_{0\to1}(t;0,0)$ does not rely on the choice of $\kappa$ because both the hazards of relapse and RRM are controlled. The directions of separable pathway effects $\SPE_{0\to2}(t;1,0)$ and $\SPE_{2\to3}(t;1,1)$ remain the same in the sensitivity analysis by varying $\kappa \in [0,1]$. The third row of Figure \ref{data} displays the estimated counterfactual cumulative incidence of mortality in the general case by the EIF-based estimator. The point estimates are similar with those in the first row. 

Table \ref{tab:test} summarizes the results of hypothesis tests on the total effect and separable pathway effects. The $p$-values of the tests of the separable pathway effects through NRM, relapse and RRM are 0.6251, 0.0105 and 0.3967 respectively under covariates isolation. As for the transition rates, significant difference between the hazards of relapse is found, but the treatment effects through NRM and RRM are insignificant. However, if we use intention-to-treat analysis to test the total effect, we would only obtain a $p$-value of 0.0186. The test given by the statistic \eqref{testsu} in the general case puts higher weight on the early phase, so the resulting $p$-value is larger when testing the separable pathway effect on relapse. Since the sample size is small and the EIF-based estimator involves many fitted models, the EIF-based estimator may have extra uncertainty compared with the GNAIPW estimator.  In summary, the proposed methods are powerful to detect the target of treatment effects from the total effect.

\begin{table}
    \centering
    \caption{Some hypothesis tests in the leukemia data, with $p$-values (CI for covariates isolation, GC for general case)}
    \begin{tabular}{lp{0.7\textwidth}cc}
    \toprule
    Test & Interpretation & $p$ CI & $p$ GC \\
    \midrule
    Total & The total treatment effect on mortality & 0.0186 & 0.4401 \\
    $\SPE_{0\to1}$ & The treatment effect on transition rates from transplantation to NRM, i.e., the separable pathway effect via NRM & 0.6251 & 0.4907 \\
    $\SPE_{0\to2}$ & The treatment effect on transition rates from transplantation to relapse, i.e., the separable pathway effect via relapse & 0.0105 & 0.0899 \\
    $\SPE_{2\to3}$ & The treatment effect on transition rates from relapse to RRM, i.e., the separable pathway effect via RRM (assuming semi-Markov) & 0.3967 & 0.8020 \\
    $\SPE_{0\to3}$ & The treatment effect on transition rates from transplantation to RRM, i.e., the separable pathway effect via relapse and RRM & 0.0032 & 0.0937 \\
    \bottomrule
    \end{tabular}
    \label{tab:test}
\end{table}

To conclude, we find that Haplo-SCT lowers the overall mortality by reducing the risk of relapse compared with MSDT. This result casts light on future guidance on allogeneic stem cell transplantation. Haplo-SCT has stronger graft-versus-leukemia effect by eradicating MRD. Since Haplo-SCT is more accessible than MSDT, it is promising that Haplo-SCT serves as an alternative to MSDT. Since the immune reconstitution undergoing Haplo-SCT is delayed due to the usage of ATG, more attention should be paid to preventing infection after transplantation. More results of the real-data analysis are provided in Supplementary Material I, including sensitivity analysis and the estimation of separable pathway effects under the general case without assuming covariates isolation. 

\section{Discussion} \label{sec8}

In this paper, we studied the identification and estimation of the counterfactual cumulative incidence of the primary (terminal) event when there is an intermediate (non-terminal) event. Under covariates isolation, baseline covariates are incorporated in the hazards by inverse probability weighting. Only a one-dimensional propensity score is needed for consistent estimation, which avoids the individual-level modeling on hazards. We further defined population-level separable pathway effects in terms of counterfactual cumulative incidences under a dismissible omponents assumption. Asymptotic properties of generalized Nelson--Aalen estimators by inverse probability weighting are studied. Confidence intervals and hypothesis testings are available for the counterfactual cumulative incidences and separable pathway effects. When covariates isolation fails but the conditional dismissible components assumption holds, we provide a efficient influence function based estimator, which enjoys asymptotic efficiency and multiple robustness. The EIF-based estimator involves more models and bears higher computational complexity. The concept of separable pathway effects provides an opportunity to understand the causal mechanism of treatment effects on the terminal event.

We paid special attention to two cases of the dependence between the intermediate and indirect outcome events, namely Markovness and semi-Markovness. Either case simplifies the dependence of $d\Lambda_3^{a_3}(t;r)$ on history information. Under dismissible components and Markovness (so that biased sampling issue will not be encountered), our work can be extended to general multi-state models. All transitions between states would be naturally aligned, and thus counterfactual incidences of all states by intervening population-level transition hazards can be generated. However, the real situation may be between Markovness and semi-Markovness. We propose a sensitivity analysis to assess the influence of Markovness or semi-Markovness with a sensitivity parameter $\kappa$. It is difficult to nonparametrically estimate $d\Lambda_3^{a_3}(t;r)$ if there is a parameter $\kappa$ in the hazard model to be estimated. There are some alternatives to this assumption by imposing model restrictions, for example, using shared frailty modeling, copula or treating the history of intermediate events as covariates in hazard modeling in semi-parametric models \cite{nevo2022causal, gorfine2023shared, sun2024penalised}.

There are some limitations for the inference by inverse probability weighting. The asymptotic properties of counterfactual cumulative incidences we derived did not consider the uncertainty of estimated propensity scores. If the true propensity scores are unknown, the inference on treatment effects would slightly deviate from the nominal level by using estimated propensity scores. The additional variation resulted by the uncertainty of estimated propensity score is the variance of a weighted expectation of martingales with respect to a coarsened filter, so the additional variance can be small in practice. If the propensity score is hard to specify, information of covariates can be utilized by elaborately modeling the cause-specific hazards and hence the probability of the instant transition and at-risk set, for example, by proportional hazards or additive hazards models. Doubly robust estimators for the hazards can be constructed by combining the propensity score model and survival probability model \cite{zhang2012contrasting}. In some literature, the censoring is considered as manipulable, and the estimand is defined by manipulating the treatment components and censoring \cite{young2020causal}. The framework in this article can be generalized to accommodate intervention on censoring to prevent loss of follow-up.

The core assumption to identify the separable effects is the dismissible components condition. Transition hazards of pathways are independent because of covariates isolation. When covariates isolation fails, identification is straightforward by covariates stratification but estimation becomes challenging. We derive the efficient influence function for the estimand and propose an efficient estimator. The estimation is both theoretically and empirically difficult without Markovness. If there are post-treatment time-varying covariates, the effect of treatment components on events can be modified by these covariates. Intervening the treatment components would change the levels of post-treatment covariates in the hypothetical experiment. For identification, dismissible treatment components for post-treatment covariates are required. We consider an alternative set of assumptions to Assumption \ref{dis} and Assumption \ref{dis_pi} in Supplementary Material G, where the treatment effects on transition hazards at time $t$ are separable conditional on all information up to $t$, including post-treatment covariates and event history. We prove the identifiability of separable pathway effects and give estimators for the separable pathway effects. However, it could be difficult to determine which treatment component has direct effect on which post-treatment covariate in practice. It is still worth studying idenfication and estimation under weaker assumptions in the presence of time-varying covariates.

\section*{Acknowledgments}
We thank Prof. Yumou Qiu (Peking University) and Dr. Yingjun Chang (Peking University People's Hospital) for helpful comments.

\section*{Data availablility statement}
The data and R codes that support this finding are available as part of the online supplementary material.

\section*{Conflict of interest}
The authors declare no conflict of interest.

\section*{Funding information}
This work is supported by National Key Research and Development Program of China, Grant No.~2021YFF0901400; National Natural Science Foundation of China, Grant No.~12026606, 12226005. This work is also partly supported by Novo Nordisk A/S.

\section*{Supplementary material}
Supplementary material, including proofs of theorems, additional simulation studies, additional real-data analysis results, are available online.

\bibliographystyle{apalike}
\bibliography{ref}

\includepdf[pages=1-42]{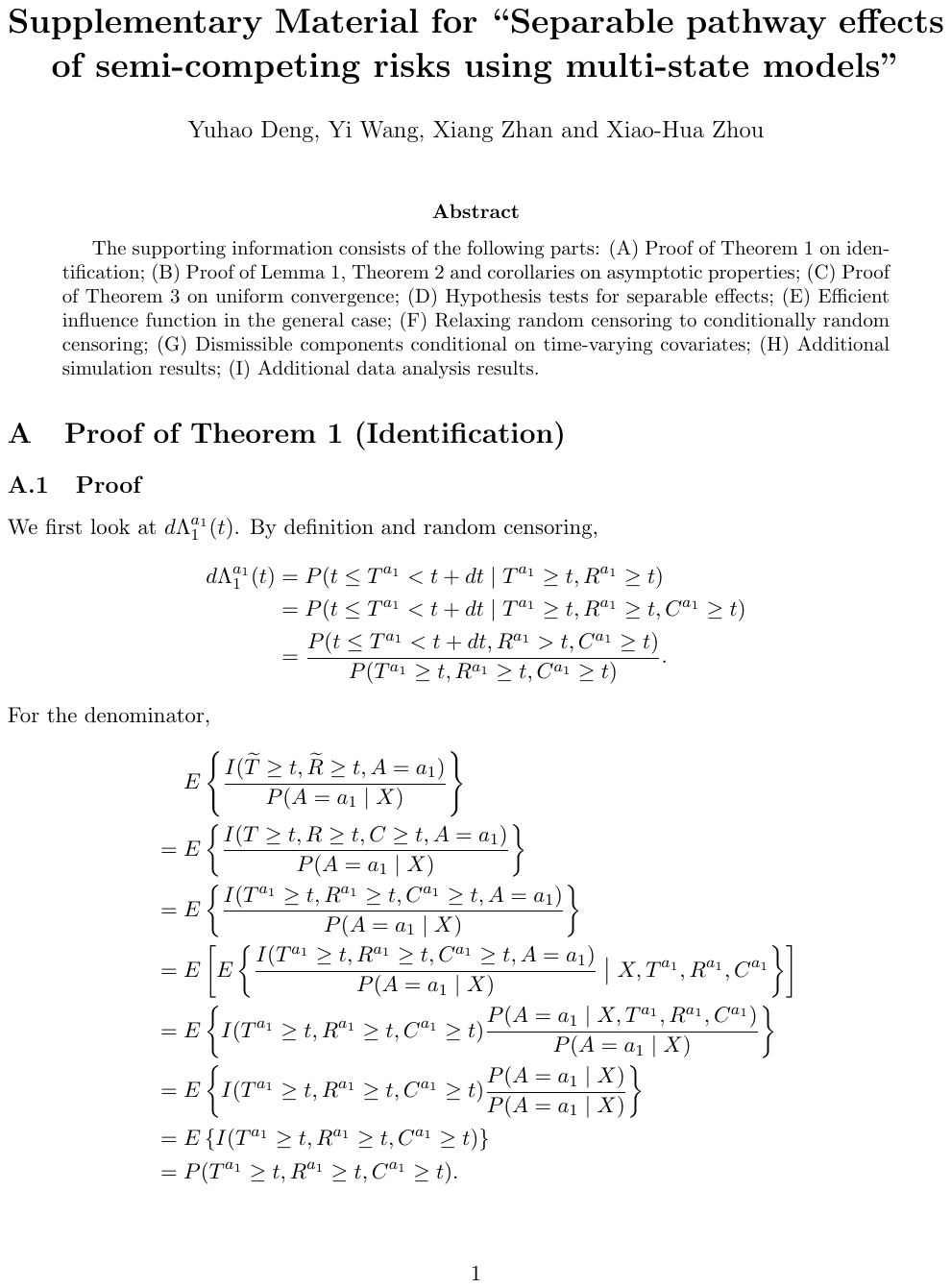}

\end{document}